\documentclass[10pt]{iopart}
\usepackage{graphicx}
\eqnobysec

\newcommand{\beq}{\begin{equation}}
\newcommand{\beqa}{\begin{eqnarray}}
\newcommand{\eeq}{\end{equation}}
\newcommand{\eeqa}{\end{eqnarray}}
\newcommand{\abs}[1]{\vert#1\vert}

\renewcommand{\bar}[1]{{\overline{#1}}}
\newcommand{\bra}[1]{\langle#1\vert}
\newcommand{\ket}[1]{\vert#1\rangle}
\newcommand{\braket}[2]{\langle#1\vert#2\rangle}
\newcommand{\braopket}[3]{\langle#1\vert#2\vert#3\rangle}
\newcommand{\dd}{{\rm d}}
\newcommand{\random}{{\,{\rm random}}}
\newcommand{\ii}{{\rm i}}
\renewcommand{\max}{{{\rm max}}}
\renewcommand{\min}{{{\rm min}}}
\newcommand{\mol}{{\rm mol}}
\newcommand{\mean}[1]{\langle#1\rangle}
\renewcommand{\tr}{\mathop{\rm tr}\,}
\newcommand{\Ai}{{\mathrm{Ai}}}
\newcommand{\Bi}{{\mathrm{Bi}}}
\renewcommand{\H}{{\cal H}}

\begin{document}

\title[Equilibration in small quantum systems: particle in 1D potential]
{An investigation of equilibration in small quantum systems:
the example of a particle in a 1D random potential}

\author{J M Luck}

\address{Institut de Physique Th\'eorique, Universit\'e Paris-Saclay, CEA and CNRS,
91191 Gif-sur-Yvette, France}

\begin{abstract}
We investigate the equilibration of a small isolated quantum system
by means of its matrix of asymptotic transition probabilities in a preferential basis.
The trace of this matrix
is shown to measure the degree of equilibration of the system
launched from a typical state,
from the standpoint of the chosen basis.
This approach is substantiated by an in-depth study
of the example of a tight-binding particle in one dimension.
In the regime of free ballistic propagation,
the above trace saturates to a finite limit, testifying good equilibration.
In the presence of a random potential,
the trace grows linearly with the system size,
testifying poor equilibration in the insulating regime induced by Anderson localization.
In the weak-disorder situation of most interest,
a universal finite-size scaling law
describes the crossover between the ballistic and localized regimes.
The associated crossover exponent 2/3 is dictated by the anomalous band-edge scaling
characterizing the most localized energy eigenstates.
\end{abstract}

\ead{\mailto{jean-marc.luck@cea.fr}}

\maketitle

\section{Introduction}

The study of equilibration and thermalization in isolated quantum systems
is as old as Quantum Mechanics itself~\cite{von,trt,gltz}.
This classic subject has experienced a complete revival in the last decade,
in parallel with the availability of new experimental results,
especially on ultra-cold atoms (see~\cite{cr,pss,efg,ge,akp} for reviews).
The matter is rather subtle.
On the one hand, an isolated quantum system undergoes unitary dynamics,
keeping thus the memory of its initial state.
On the other hand, if the system under scrutiny is large enough,
Statistical Mechanics is expected to apply, at least approximately,
when local observables and small subsystems are considered.
The latter assertion is clearly very general,
and would apply to classical systems as well.

A great deal of recent works have aimed at answering fundamental questions
concerning equilibration and thermalization of {\it large} isolated quantum
systems~\cite{psw,rdy,rdo,prf,bs,mr,lps,lp,pr,bkl,gpe,as,lep,iu,ghlt,ib}.
Most questions addressed in these works,
and tackled in many other contributions as well,
concern the relevance of statistical-mechanical concepts such as
ergodicity, temperature and ensembles,
including the effective description of a small subsystem
by a microcanonical or a canonical ensemble,
the role of conservation laws, especially in integrable systems,
and the thermalization properties of single highly-excited states.

The focus of the present work is different.
Our purpose is to have a closer look at {\it small} isolated quantum systems.
We characterize their degree of equilibration
-- or of lack of equilibration --
by the matrix of asymptotic transition probabilities
in a preferential basis chosen once for all.
In this context, by equilibration we mean
(weak) convergence to a unique stationary state,
irrespective of the initial state of the system.
The motivation for the present work stems from a recent observation
concerning a tight-binding quantum particle on a finite chain.
Whereas the probability distribution of a classical random walker
converges exponentially fast to a unique stationary state,
the stationary state of a quantum walker (modelled as a tight-binding particle)
keeps forever a weak but significant memory of its initial position
along the chain~\cite{klm}.

The plan of this paper is as follows.
In section~\ref{gal} we present some general formalism
and introduce the key objects of this work,
namely the matrix $Q$ of asymptotic transition probabilities in a preferential basis
and its trace $T$.
We then analyze in detail the example of a tight-binding particle on a finite chain.
The case of a free particle is investigated in section~\ref{free}.
The next three sections are devoted to the richer situation where the particle
is subjected to a static random potential and may experience Anderson localization.
General features of the localized regime are described in section~\ref{wgal}.
The situation of a weak disorder is investigated in section~\ref{weak}.
We show that the trace $T$ manifests universal behavior
dictated by the anomalous band-edge scaling,
in the weak-disorder localized regime
and throughout the crossover between the ballistic and localized regimes.
The non-universal features which characterize the strong-disorder regime
are investigated in section~\ref{strong}.
In section~\ref{disc} we summarize our findings,
and speculate on possible extensions of the present approach
to other situations, including many-body quantum systems.

\section{Generalities}
\label{gal}

We consider an isolated quantum system whose Hilbert space has finite dimension~$N$.
This system is endowed with a preferential basis $\ket{a}$ ($a=1,\dots,N$),
chosen on some physical grounds,
e.g.~because it is easy to prepare the system in any of the basis states~$\ket{a}$.
In that basis the Hamiltonian is an $N\times N$ Hermitian matrix~$\H$.
We assume for simplicity
that the energy eigenvalues~$E_n$ ($n=1,\dots,N$) are non-degenerate.
Let~$\ket{n}$ be normalized eigenvectors so that $\H\ket{n}=E_n\ket{n}$.

If the system is initially prepared in one of the basis states, say $\ket{a}$,
its state vector $\ket{\psi(t)}$ at subsequent times reads
\beq
\ket{\psi(t)}=\sum_n\e^{-\ii E_nt}\ket{n}\braket{n}{a}.
\eeq
The probability of observing the system is state $\ket{b}$ at time $t$ is therefore
\beq
P_{ab}(t)=\abs{\braket{b}{\psi(t)}}^2
=\sum_{m,n}\e^{\ii(E_n-E_m)t}\braket{b}{m}\braket{m}{a}\braket{a}{n}\braket{n}{b}.
\label{pab}
\eeq

Henceforth we focus our attention onto the asymptotic stationary state
reached by the system,
defined by considering the time-averaged transition probabilities
\beq
Q_{ab}=\lim_{t\to\infty}\frac{1}{t}\int_0^tP_{ab}(t')\,\dd t'.
\eeq
Because of the absence of spectral degeneracies, only the diagonal terms ($m=n$)
of the double sum in~(\ref{pab}) contribute to this limit.
We thus obtain
\beq
Q_{ab}=\sum_n\abs{\braket{a}{n}}^2\;\abs{\braket{b}{n}}^2.
\label{qab}
\eeq

The matrix $Q$ so defined is the central object of the present work.
It characterizes entirely the asymptotic stationary state
issued from the initial state $\ket{a}$.
The non-triviality of $Q$ reflects the fact that
an isolated quantum system undergoing unitary dynamics
remembers its initial state forever.
This everlasting memory affects
asymptotic time-averaged quantities such as $Q_{ab}$.
Let us start with a few general properties.
The matrix $Q$ is real symmetric and positive definite, as it reads
\beq
Q=RR^T,
\eeq
where the matrix $R$ is defined by
\beq
R_{an}=\abs{\braket{a}{n}}^2,
\label{rdef}
\eeq
and $R^T$ denotes the transpose of $R$.
The matrices $Q$ and $R$ are doubly stochastic,
i.e., their row and column sums equal unity:
\beq
\sum_aR_{an}=\sum_nR_{an}=\sum_aQ_{ab}=\sum_bQ_{ab}=1.
\eeq
The spectra of $Q$ and $R$ are however not simply related to each other in general,
as the matrix $R$ is not symmetric and may therefore have complex spectrum.

An alternative presentation consists in introducing the density matrix
\beq
\rho_a(t)=\ket{\psi(t)}\bra{\psi(t)}
=\sum_{m,n}\e^{\ii(E_n-E_m)t}\ket{m}\braket{m}{a}\braket{a}{n}\bra{n},
\eeq
whose time-averaged value
\beq
\omega_a=\lim_{t\to\infty}\frac{1}{t}\int_0^t\rho_a(t')\,\dd t'
\eeq
reads
\beq
\omega_a=\sum_n\abs{\braket{a}{n}}^2\;\ket{n}\bra{n}.
\label{oares}
\eeq
The quantities $P_{ab}(t)$ and $Q_{ab}$ can be recovered as
\beq
P_{ab}(t)=\tr\rho_a(t)\Pi_b,\qquad Q_{ab}=\tr\omega_a\Pi_b,
\eeq
where $\Pi_b=\ket{b}\bra{b}$ is the projector onto the final state.

The asymptotic transition probabilities have the alternative expression
\beq
Q_{ab}=\tr\omega_a\omega_b.
\eeq
The diagonal elements $Q_{aa}$ of the matrix $Q$ therefore represent two things.
First, by definition,
\beq
Q_{aa}=\sum_n\abs{\braket{a}{n}}^4
\label{qaa1}
\eeq
is the stationary return probability to state $\ket{a}$.
Second,
\beq
Q_{aa}=\tr\omega_a^2
\label{qaa2}
\eeq
is the so-called purity~\cite{qmafp,qmba}
of the stationary density matrix~$\omega_a$ issued from state~$\ket{a}$.
Its reciprocal
\beq
d_a=\frac{1}{Q_{aa}}=\frac{1}{\tr\omega_a^2}
\eeq
is the effective number of eigenstates~$\ket{n}$ involved in $\omega_a$,
i.e., the effective dimension of the subspace
in which the system equilibrates if launched from state~$\ket{a}$.
This interpretation has already been put forward in~\cite{prf,lps,lp,gpe,as,iu}.

Our main goal is to investigate the quantum evolution
from a typical initial basis state.
In this context, it is natural to consider the trace of the matrix $Q$,
\beq
T=\tr Q=\sum_{a,n}\abs{\braket{a}{n}}^4,
\label{tdef}
\eeq
and to interpret the ratio
\beq
D=\frac{N}{T}
\label{ddef}
\eeq
as the effective dimension of the subspace
which hosts the equilibration dynamics of a typical initial state.
We have indeed
\beq
\frac{1}{D}=\frac{1}{N}\sum_a\frac{1}{d_a}.
\eeq

The trace $T$ can alternatively be viewed as the sum
of the inverse participation ratios
$I_n$ of all eigenstates $\ket{n}$ (see~(\ref{iprdef}),~(\ref{tsum})).

The trace $T$ and the effective dimension $D$ always sit
between extremal values corresponding to the following two limiting situations.

\subsubsection*{Ideal equilibration.}
This situation is met when the energy eigenstates
are uniformly spread in the preferential basis.
We have
\beq
R_{an}=\abs{\braket{a}{n}}^2=\frac{1}{N}
\eeq
for all $a$ and $n$, and so
\beq
Q_{ab}=\frac{1}{N}
\eeq
for all $a$ and $b$:
the asymptotic transition probability $Q_{ab}$
is independent of the initial state $\ket{a}$ and on the final state $\ket{b}$.
This situation therefore corresponds to ideal equilibration.
The trace assumes its minimal value
\beq
T_\min=1\qquad(D_\max=N).
\label{tmin}
\eeq

\subsubsection*{No equilibration.}
This situation is the exact opposite of the previous one.
It occurs when the Hamiltonian $\H$ is diagonal in the preferential basis,
so that the energy eigenstates $\ket{n}$ coincide with the basis states $\ket{a}$.
We can therefore reorder the eigenstates $\ket{n}$ so as to have
\beq
R_{an}=\abs{\braket{a}{n}}^2=\delta_{a,n}
\eeq
for all $a$ and $n$, and so
\beq
Q_{ab}=\delta_{a,b}
\eeq
for all $a$ and $b$.
In such a situation, the system stays in its initial state forever:
there is no equilibration at all.
The trace assumes its maximal value
\beq
T_\max=N\qquad(D_\min=1).
\label{tmax}
\eeq

\medskip

In view of the above,
the behavior of the trace $T$ as a function of the dimension~$N$ of the Hilbert space
is expected to characterize
the degree of equilibration of an isolated quantum system,
from the viewpoint of the chosen preferential basis.
If $T$ is much smaller than~$N$ (i.e.,~$D$ is almost as large as~$N$),
typical energy eigenstates are rather extended in that basis,
and so the system will show good equilibration.
Conversely, if $T$ is comparable to $N$ (i.e., $D$ is much smaller than~$N$),
typical energy eigenstates are rather localized in that basis,
and so the system will keep a strong memory of its initial state
and hardly exhibit any equilibration.
These statements extend the ideas presented in~\cite{prf,lps,lp,gpe,as,iu}
and generalize them to a typical initial state,
from the standpoint of the chosen preferential basis.

In the next sections, we shall substantiate the above formalism
by an in-depth study of the example of a tight-binding particle in one dimension,
both in the regime of free propagation and in the presence of a random potential.

To close this general section,
let us discuss how the above formalism
is modified by the presence of spectral degeneracies.
We denote by $\mu_n$ the multiplicity (degeneracy) of the $n$th energy eigenvalue~$E_n$,
with
\beq
\sum_n\mu_n=N,
\eeq
and introduce an orthonormal basis of vectors $\ket{n,i}$ such that
$\H\ket{n,i}=E_n\ket{n,i}$ ($i=1,\dots,\mu_n$).
The orthogonal projector $\Pi_n$ onto the eigenspace associated with~$E_n$
has rank $\mu_n$ and reads
\beq
\Pi_n=\sum_i\ket{n,i}\bra{n,i}.
\eeq
The expression~(\ref{qab}) for the asymptotic transition probability $Q_{ab}$ becomes
\beqa
Q_{ab}&=&\sum_n\tr(\Pi_a\Pi_n\Pi_b\Pi_n)
\nonumber\\
&=&\sum_n\sum_{i,j}
\braket{a}{n,i}\braket{a}{n,j}^*\,\braket{b}{n,j}\braket{b}{n,i}^*,
\label{dqab}
\eeqa
where the star denotes complex conjugation.
In particular, the expressions~(\ref{qaa1})
for the stationary return probability $Q_{aa}$
and~(\ref{tdef}) for the trace $T$ become
\beq
Q_{aa}=\sum_n\biggl(\sum_i\abs{\braket{a}{n,i}}^2\biggr)^2,
\label{dqaa1}
\eeq
\beq
T=\sum_{a,n}\biggl(\sum_i\abs{\braket{a}{n,i}}^2\biggr)^2.
\label{dtdef}
\eeq

\section{Free particle}
\label{free}

In this section we investigate the simple situation
of a free tight-binding particle on a finite chain of $N$ sites.
This problem has already been tackled in~\cite[Appendix~A]{klm}.
The preferential basis consists of the local Wannier states $\ket{a}$ ($a=1,\dots,N$).
Introducing the amplitudes $\psi_a=\braket{a}{\psi}$,
we write down the eigenvalue equation $\H\psi=E\psi$
as
\beq
E\psi_a=\psi_{a+1}+\psi_{a-1}.
\eeq

For an open chain with Dirichlet boundary conditions ($\psi_0=\psi_{N+1}=0$),
the non-degenerate energy eigenvalues and eigenstates read
\beqa
E_n=2\cos\frac{n\pi}{N+1},
\label{en}
\\
\braket{a}{n}=\sqrt\frac{2}{N+1}\sin\frac{an\pi}{N+1},
\label{an}
\eeqa
($n=1,\dots,N$), and so
\beq
R_{an}=\frac{2}{N+1}\,\sin^2\frac{an\pi}{N+1}.
\label{rfree}
\eeq
This situation is one of the rare cases where the matrix $R$ is symmetric.
We have
\beq
Q_{ab}=\left(\frac{2}{N+1}\right)^2\sum_n\sin^2\frac{an\pi}{N+1}\sin^2\frac{bn\pi}{N+1}.
\eeq
This sum can be worked out explicitly by expanding the sines into complex exponentials
and using the identity
\beq
\sum_{n=1}^N\e^{2\ii cn\pi/(N+1)}=(N+1)\delta_{c,0}-1,
\label{iden}
\eeq
where the Kronecker function acts modulo $N+1$:
$\delta_{c,0}$ is non-zero if and only if the integer $c$ is a multiple of $N+1$.
We thus obtain
\beq
Q_{ab}=\frac{1}{N+1}\left(1+\frac{1}{2}\,\delta_{a,b}+\frac{1}{2}\,\delta_{a+b,N+1}\right).
\eeq
With respect to its background value $1/(N+1)$,
the transition probability~$Q_{ab}$ is enhanced by a factor 3/2
at the starting point ($b=a$) and at the mirror-symmetric position ($b=N+1-a$).
In the special case where the starting point is at the middle of an odd chain
($N$ odd and $a=(N+1)/2$), the enhancement factor of the return probability reaches 2.

The corresponding trace
\beq
T=\frac{6N+1-(-1)^N}{4(N+1)}
\label{tfree}
\eeq
saturates to the finite value
\beq
T=\frac{3}{2}
\label{tasyfree}
\eeq
for large system sizes.

This limit is larger than the minimal value~(\ref{tmin}) by a finite factor.
Our approach thus predicts that a free (i.e., ballistic) quantum particle
on a finite system has good, albeit not ideal, equilibration properties.

This conclusion can be corroborated
by looking at the position $X$ of a particle launched from site~$a$.
The stationary mean value of $X$,
\beq
\mean{X}=\tr(\omega_aX)=\sum_b bQ_{ab}=\frac{N+1}{2},
\eeq
is dictated by the symmetry of the energy eigenstates,
and therefore strictly independent of the initial state.
The mean square position reads
\beqa
\mean{X^2}&=&\tr(\omega_aX^2)=\sum_b b^2Q_{ab}
\nonumber\\
&=&\frac{(N+1)^2}{3}-\frac{a(N+1-a)}{N+1}+\frac{1}{6}.
\eeqa
The second term exhibits a weak dependence, of relative order $1/N$,
on the particle's initial position~$a$.

\section{Particle in a random potential: generalities}
\label{wgal}

We now address the richer situation where the quantum particle
is subjected to a static random potential.
The energy eigenvalue equation now reads
\beq
E\psi_a=\psi_{a+1}+\psi_{a-1}+V_a\psi_a.
\label{hamv}
\eeq
We still consider an open chain of $N$ sites with Dirichlet boundary conditions
($\psi_0=\psi_{N+1}=0$).
The on-site energies~$V_a$ which build up the random potential
are independent from each other and drawn from a symmetric distribution $f(V)$,
so that\footnote{A bar denotes an average
with respect to the distribution $f(V)$ of the random potential.}
$\bar{V}=0$ and $\bar{V^2}=w^2$,
where $w$ is referred to as the strength of the disorder.

The symmetric uniform and binary distributions
\beqa
f_{\rm uni}(V)&=&\frac{1}{2w\sqrt3}\qquad(-w\sqrt3<V<w\sqrt3),
\label{uni}
\\
f_{\rm bin}(V)&=&\frac{1}{2}\bigl(\delta(V-w)+\delta(V+w)\bigr)
\label{bin}
\eeqa
will be used hereafter in numerical studies.

It is a well-known feature of Anderson localization that
all eigenstates are exponentially localized in one dimension,
irrespective of the disorder strength~$w$~\cite{gang,lgp,fifty}.
In the weak-disorder regime ($w\to0$),
the spectrum is close to that of a free particle,
i.e., a band parametrized by the dispersion relation
\beq
E=2\cos q.
\eeq
Within the band, the localization length diverges according to the universal law
\beq
\xi\approx\frac{8\sin^2q}{w^2},
\label{xipert}
\eeq
known as the Thouless formula~\cite{thh}.
This perturbative estimate breaks down near the band edges
($E\to\pm2$, i.e., $q\to0$ and $\pi$), where the numerator vanishes.
Right at the band edges, eigenstates are actually more strongly localized,
as their localization length only diverges as $w^{-2/3}$~\cite{flo,hal,dgedge,irt,cltt}.
This anomalous band-edge scaling will play a key role in the following
(see section~\ref{weak}).
For a larger disorder strength $w$, beyond the above universal regime,
all energy eigenstates are strongly localized,
as the localization length becomes comparable to the lattice spacing.

The degree of localization of eigenstate $\ket{n}$
is commonly characterized by the inverse participation ratio
(IPR)~\cite{bdh,bd,vis,th}
\beq
I_n=\sum_a\braket{a}{n}^4,
\label{iprdef}
\eeq
where $\psi_{a,n}=\braket{a}{n}$ is the normalized real wavefunction
of eigenstate $\ket{n}$ at site $a$.
The participation ratio (PR)
\beq
\ell_n=\frac{1}{I_n}
\label{prdef}
\eeq
measures how many sites `participate' in eigenstate~$\ket{n}$,
in the sense that their weights $\braket{a}{n}^2$ are appreciable.
It therefore gives an estimate of the spatial extent of that eigenstate.

The definitions~(\ref{qaa1}) of the return probabilities $Q_{aa}$
and~(\ref{iprdef}) of the IPR $I_n$ are dual to each other:
$d_a=1/Q_{aa}$ measures how many eigenstates~$\ket{n}$
have appreciable weights at site $a$,
while $\ell_n=1/I_n$ measures how many sites $a$
have appreciable weights in eigenstate~$\ket{n}$.
We have in particular the sum rules
\beq
T=\sum_aQ_{aa}=\sum_nI_n.
\label{tsum}
\eeq

In the weak-disorder regime,
the spatial extent $\ell_n=1/I_n$ of typical eigenstates
is expected to be as large as the localization length,
and therefore to diverge according to the Thouless formula~(\ref{xipert}).
As soon as disorder is strong enough,
typical values of~$\ell_n$ becomes much smaller than the system size $N$.
In this insulating regime,
it is to be expected that the return probabilities $Q_{aa}$ and the IPR $I_n$
are fluctuating quantities of the same order of magnitude,
whose distributions become asymptotically independent of $N$.
In particular, their common mean value
\beq
\bar{Q}=\bar{I}
\label{qiiden}
\eeq
dictates the asymptotic linear growth of the trace $T$ with the system size:
\beq
T\approx\bar{Q}N.
\label{tdis}
\eeq

The trace therefore grows proportionally to its maximal value~(\ref{tmax}),
albeit~with a smaller amplitude $\bar{Q}<1$.
This key result is to be contrasted with its counterpart~(\ref{tasyfree})
in the case of a free particle.
Whereas a quantum particle has good equilibration properties in the regime
of free ballistic propagation,
it has poor equilibration properties in the insulating or localized regime.
Indeed, as a consequence of~(\ref{ddef}) and~(\ref{tdis}),
its equilibration dynamics typically takes place in a subspace whose dimension
\beq
D=1/\bar{Q}
\eeq
stays finite in the limit of an infinite chain.

The next two sections are devoted to a quantitative analysis of the problem
in the weak-disorder regime (section~\ref{weak})
and in the strong-disorder regime (section~\ref{strong}).
Before this, it is worth illustrating the above concepts and results
by means of numerical results for a generic disorder strength.

\begin{figure}[!ht]
\begin{center}
\includegraphics[angle=-90,width=.5\linewidth]{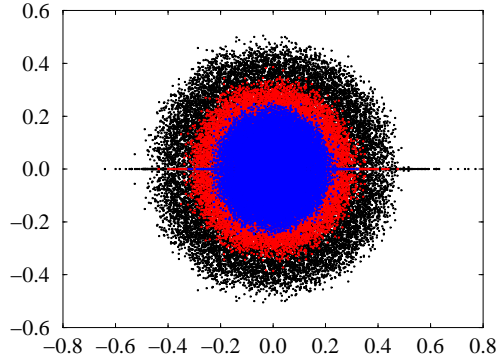}
\caption{\small
Complex spectrum of the matrix $R$
for a uniform distribution of the random potential
with three relatively small disorder strengths:
$w=0.1$ (blue), $w=0.3$ (red) and $w=1$ (black).}
\label{rcomplex}
\end{center}
\end{figure}

We begin with the matrix $R$ defined in~(\ref{rdef}).
In the case of a free particle, this matrix is symmetric (see~(\ref{rfree})).
As soon as a random potential is present, this symmetry property is broken,
and we are facing the generic situation where $R$ has complex spectrum.
This is pictured in figure~\ref{rcomplex}, showing this spectrum
for a uniform distribution of the potential with three different
disorder strengths $w$.
For each $w$, the spectra of 300 samples of size $N=100$ are superimposed.
The spectra appear to be roughly circular, with a radius growing slowly with $w$.

\begin{figure}[!ht]
\begin{center}
\includegraphics[angle=-90,width=.5\linewidth]{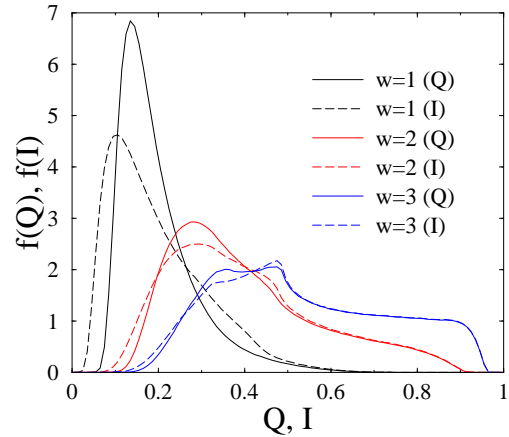}
\caption{\small
Distributions $f(Q)$ of the return probabilities $Q_{aa}$ (full curves)
and $f(I)$ of the IPR $I_n$ (dashed curves)
for a uniform distribution of the random potential
with three different disorder strengths $w$.}
\label{qi}
\end{center}
\end{figure}

We then look at the distributions $f(Q)$
of the return probabilities $Q_{aa}$ and $f(I)$ of the IPR $I_n$.
Figure~\ref{qi} shows plots of these probability distributions,
for a uniform distribution of the random potential with three different
disorder strengths $w$.
For each value of $w$, data are gathered over $10^5$ samples of size $N=100$.
Finite-size effects are negligible.
As expected from the above discussion,
the distributions $f(Q)$ (full curves) and $f(I)$ (dashed curves)
are very similar to each other.
The distribution of the IPR is only slightly broader, especially at weak disorder.

\begin{figure}[!ht]
\begin{center}
\includegraphics[angle=-90,width=.5\linewidth]{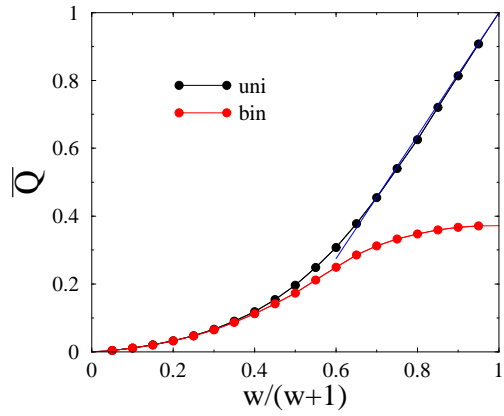}
\caption{\small
Mean return probability $\bar{Q}$ against $w/(w+1)$
for uniform and binary distributions of the random potential.
Blue straight line: strong-disorder estimate~(\ref{qstrong}),~(\ref{qsuni})
for the uniform distribution.}
\label{qave}
\end{center}
\end{figure}

We finally investigate the mean return probability~$\bar{Q}$,
which is the key quantity governing the asymptotic linear growth~(\ref{tdis})
of the trace $T$.
The overall dependence of $\bar{Q}$ on the type and strength of disorder
is illustrated in figure~\ref{qave}, showing $\bar{Q}$ against $w/(w+1)$
for uniform and binary distributions of the random potential.
The choice of abscissa allows a convenient representation
of the infinite-disorder limit.
For the system size $N=400$ used here, finite-size corrections
are much smaller than statistical errors,
which are themselves smaller than the symbol size.
For both types of disorder,
the mean return probability increases smoothly with the disorder strength $w$.
It will be shown in section~\ref{weak} that $\bar{Q}$ obeys a universal growth in $w^{4/3}$
in the weak-disorder regime (see~(\ref{qtweak}),~(\ref{a21})).
Its behavior in the strong-disorder regime depends on the type of disorder.
This question will be studied in section~\ref{strong}.
For a continuous distribution, such as the uniform one,~$\bar{Q}$ goes to unity
with a finite slope, i.e., a leading $1/w$ correction (see~(\ref{qstrong})).
Discrete distributions give rise to a non-trivial infinite-disorder limit,
which reads $\bar{Q}_\infty\approx0.373$ for the binary distribution (see~(\ref{qbin})).

\section{Particle in a random potential: weak-disorder regime}
\label{weak}

This section is devoted to an in-depth analysis of the weak-disorder regime.
Our main goals are to determine the behavior of the key quantity $\bar{Q}$ in this regime,
and to investigate the crossover between the laws~(\ref{tasyfree}) and~(\ref{tdis}).

\subsection{Perturbation theory}
\label{weakpert}

Let us begin by deriving the prediction of standard second-order perturbation theory
for the trace $T$.
In the presence of the random potential, the unnormalized $n$-th eigenstate reads
\beqa
\ket{n}_\random=\ket{n}&+&\sum_{m\ne n}\frac{\braopket{m}{V}{n}}{E_n-E_m}\ket{m}
\nonumber\\
&+&\sum_{m\ne n}\sum_{l\ne n}\frac{\braopket{m}{V}{l}\braopket{l}{V}{n}}
{(E_n-E_m)(E_n-E_l)}\ket{m}
\nonumber\\
&-&\braopket{n}{V}{n}\sum_{m\ne n}\frac{\braopket{m}{V}{n}}{(E_n-E_m)^2}\ket{m}+\cdots,
\label{nran}
\eeqa
where $E_n$ and $\ket{n}$ pertain to the free problem (see~(\ref{en}),~(\ref{an})).

An expansion of the disorder-averaged trace $T$
can be derived by inserting the expression~(\ref{nran}),
once properly normalized, into the definition~(\ref{tdef}).
Using
\beq
\bar{\braopket{k}{V}{l}\braopket{m}{V}{n}}
=w^2\sum_a\braket{a}{k}\braket{a}{l}\braket{a}{m}\braket{a}{n},
\eeq
we obtain
\beq
T=T^{(0)}+T^{(2)}w^2+\cdots,
\label{tpert}
\eeq
where $T^{(0)}$ is given by~(\ref{tfree}), while
\beqa
T^{(2)}=\sum_{a,b,n}
&&\Big(4\braket{a}{n}\braket{b}{n}\bigl(\braket{a}{n}^2-I_n\bigr)
\bigl(X_{a b n}X_{b b n}-\braket{b}{n}^2Y_{a b n}\bigr)
\nonumber\\
&&+(6\braket{a}{n}^2-2I_n\bigr)\braket{b}{n}^2X_{a b n}^2\Big),
\label{t2full}
\eeqa
with the notations~(\ref{en}),~(\ref{an}),~(\ref{iprdef}) and the shorthands
\beq
X_{a b n}=\sum_{m\ne n}\frac{\braket{a}{m}\braket{b}{m}}{E_n-E_m},\qquad
Y_{a b n}=\sum_{m\ne n}\frac{\braket{a}{m}\braket{b}{m}}{(E_n-E_m)^2}.
\eeq
The quantities $T^{(0)}$ and $T^{(2)}$ are rational numbers for any system size $N$.
Table~\ref{pert} gives their explicit values for $N$ up to 10.

\begin{table}[!ht]
\begin{center}
\begin{tabular}{|c|c|c|c|c|c|c|c|c|c|c|}
\hline
$N$&1&2&3&4&5&6&7&8&9&10\\
\hline
$T^{(0)}$&$1$&$1$&$\frac{5}{4}$&$\frac{6}{5}$&$\frac{4}{3}$&$\frac{9}{7}$&$\frac{11}{8}$&$\frac{4}{3}$&$\frac{7}{5}$&$\frac{15}{11}$\\
\hline
$T^{(2)}$&$0$&$\frac{1}{2}$&$\frac{1}{64}$&$\frac{243}{250}$&$\frac{5}{9}$&$\frac{71}{49}$&$\frac{325}{256}$&$\frac{61}{27}$&$\frac{517}{250}$&$\frac{787}{242}$\\
\hline
\end{tabular}
\caption{Exact rational values of the terms $T^{(0)}$ and $T^{(2)}$
of the expansion~(\ref{tpert}) of the disorder-averaged trace $T$,
for system sizes $N$ up to 10.}
\label{pert}
\end{center}
\end{table}

The triple sum entering~(\ref{t2full})
can be worked out explicitly as a function of $N$
by means of the repeated use of identities such as~(\ref{iden})
and generalizations thereof,
with the help of the symbolic software MACSYMA.
We prefer to skip every detail and just give the result
\beq
T^{(2)}=\frac{13(N+1)^2}{540}+\frac{\Delta_N}{4320(N+1)^2},
\eeq
with
\beqa
\Delta_N&=&181+675(2N^2+4N+1)(-1)^N
\nonumber\\
&+&320\left(\frac{4N+7}{\sqrt3}\sin\frac{2N\pi}{3}-(4N+3)\cos\frac{2N\pi}{3}\right).
\eeqa
This formula testifies the complexity of the problem at hand.
The quantity~$T^{(0)}$ (see~(\ref{tfree}))
contains a damped oscillatory term in $(-1)^N$ with period 2 in $N$,
while~$T^{(2)}$ has damped oscillatory terms with periods 2 and 3.

For a long sample,
and more precisely in the regime where $w\to0$ before $N\to\infty$,
the expansion~(\ref{tpert}) simplifies to
\beq
T\approx\frac{3}{2}+\frac{13}{540}N^2w^2.
\label{twn}
\eeq

\subsection{Weak-disorder localized regime}
\label{locweak}

Let us now consider the weak-disorder localized regime,
corresponding to the opposite order of limits, i.e., $N\to\infty$ before $w\to0$.
The second sum in~(\ref{tsum}) expresses the trace $T$
as a sum of the IPR of all the energy eigenstates.
In the weak-disorder regime ($w\to0$),
the IPR $I_n\sim1/\ell_n$ of generic band states scales as $w^2$,
namely as the reciprocal of the perturbative estimate~(\ref{xipert})
for the localization length.
Edge states, whose energies are close to band edges ($E_n\approx\pm2$),
are however much more strongly localized, and their IPR is expected to scale as $w^{2/3}$.

In order to build up an estimate for the trace $T$,
we need to know how many such edge states contribute to~(\ref{tsum}).
This number can be estimated by making use of anomalous band-edge scaling,
which has been investigated first in continuum models
with a white-noise potential~\cite{flo,hal},
and then at the band edges of tight-binding models~\cite{dgedge,irt,cltt}.
The main result of the latter works can be expressed in the following compact form.
The complex characteristic exponent $\Omega(E)=\gamma(E)+\ii\pi H(E)$,
where $\gamma(E)=1/\xi(E)$ is the reciprocal of the localization length,
while $H(E)$ is the integrated density of states (between $E$ and~$+\infty$),
obeys the scaling law
\beq
\Omega(E)\approx(w^2/2)^{1/3}\;G\!\left(\frac{E-2}{(w^2/2)^{2/3}}\right)
\label{Osca}
\eeq
at the upper band edge ($E\to2$),
throughout the regime where $w$ and $E-2$ are simultaneously small.
The complex scaling function
\beq
G(x)=\e^{-2\ii\pi/3}\frac{\Ai'(\e^{-2\ii\pi/3}x)}{\Ai(\e^{-2\ii\pi/3}x)}
=\frac{\Ai'(x)+\ii\,\Bi'(x)}{\Ai(x)+\ii\,\Bi(x)}
\eeq
involves Airy functions $\Ai$ and $\Bi$.

The scaling law~(\ref{Osca}) implies that the inverse localization length
$\gamma(E)$ and the integrated density of states $H(E)$ scale as $w^{2/3}$
in a scaling region of width $\Delta E\sim w^{4/3}$ around
each band edge ($E\approx\pm2$).
Coming back to the trace~$T$,
the second sum in~(\ref{tsum}) involves $N$ band states, whose IPR scales as $w^2$,
and a number of order $Nw^{2/3}$ of edge states, whose IPR scales as $w^{2/3}$.
Edge states therefore govern the behavior of $\bar{Q}$ and of $T$ at weak disorder,
which reads
\beq
\bar{Q}\approx Aw^{4/3},\qquad
T\approx Aw^{4/3}N.
\label{qtweak}
\eeq

We have no analytical prediction for the amplitude $A$.
Deriving such a prediction would require an extension
to the full band-edge scaling regime
of the analysis of the IPR distribution at weak disorder performed in~\cite{ky}.
This task looks formidable.
Investigating the distribution of the IPR is also known
to be difficult in other contexts,
as testified by a recent study of this question
for the discrete Laplacian on random regular graphs~\cite{cm}.

\begin{figure}[!ht]
\begin{center}
\includegraphics[angle=-90,width=.5\linewidth]{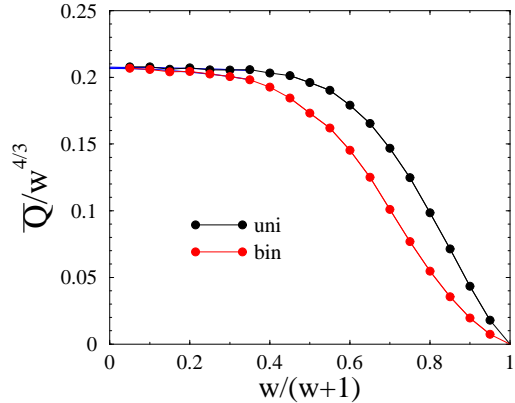}
\caption{\small
Ratio $\bar{Q}/w^{4/3}$ against $w/(w+1)$,
for uniform and binary distributions of the random potential.
Data for $\bar{Q}$ are taken from figure~\ref{qave}.
Blue curves: quadratic fits of the first 7 points of each dataset
yielding $A\approx0.207$ and $A\approx0.206$.}
\label{qaveter}
\end{center}
\end{figure}

In order to obtain a numerical value of the amplitude $A$,
we have replotted the data of figure~\ref{qave} in figure~\ref{qaveter},
where the ratio $\bar{Q}/w^{4/3}$ is shown against $w/(w+1)$.
The data for uniform and binary disorder
have a very weak dependence on $w$ in the weak-disorder regime.
Quadratic fits of the first 7 points of each series (hardly visible blue curves)
respectively yield $A\approx0.207$ and $A\approx0.206$.
We thus obtain an accurate determination of the amplitude
\beq
A\approx0.21.
\label{a21}
\eeq

\subsection{Weak-disorder crossover between ballistic and localized regimes}
\label{weakxover}

The aim of this section is to study the full weak-disorder crossover
between the ballistic and localized regimes,
which takes place when $w\to0$ and $N\to\infty$ simultaneously.
The line of thought of section~\ref{locweak} can be summarized as follows.
In the weak-disorder regime,
the second sum entering the expression~(\ref{tsum}) of the trace $T$
is dominated by edge states.
The number of these edge states grows as $Nw^{2/3}$,
while the PR of each of them scales as~$w^{2/3}$.
These estimates give rise to~(\ref{qtweak})
as long as $N$ is large enough, i.e., much larger than $w^{-2/3}$.

Two phenomena occur simultaneously at the crossover length $N\sim w^{-2/3}$:
the number of edge states becomes of order unity,
while their localization length becomes comparable to the system size.
Putting everything together,
we are led to write down a universal finite-size scaling law of the form
\beq
T\approx T^{(0)}+\frac{1}{N}\,\Phi\bigl(Nw^{2/3}\bigr).
\label{tfss}
\eeq
The first term $T^{(0)}$ is the value of $T$ in the absence of disorder,
given by~(\ref{tfree}), and so we have $\Phi(0)=0$.
The prefactor $1/N$ of the second term
testifies that the number of relevant edge states
becomes of order unity at the crossover length.

Surprisingly enough,
if the limiting value $3/2$ (see~(\ref{tasyfree}))
had been chosen as the first term of~(\ref{tfss})
instead of the exact $N$-dependent expression~(\ref{tfree}) for $T^{(0)}$,
the finite-size scaling function $\Phi(x)$ would have been corrected
by a finite offset depending on the parity of the system size,
i.e., $\Phi(0)=-3/2$ (resp.~$\Phi(0)=-1$) for $N$ even (resp.~$N$ odd).

The finite-size scaling formula~(\ref{tfss})
interpolates between~(\ref{tasyfree}) and~(\ref{qtweak}),
respectively corresponding to the ballistic and localized regimes.
When $x=Nw^{2/3}$ is small,
the results~(\ref{tasyfree}) and~(\ref{twn}) are recovered as
\beq
\Phi(x)\approx\frac{13\,x^3}{540}\qquad(x\to0).
\label{phis}
\eeq
When $x=Nw^{2/3}$ is large,
the linear growth~(\ref{qtweak}) is recovered as
\beq
\Phi(x)\approx A\,x^2\qquad(x\to\infty).
\label{phil}
\eeq

\begin{figure}[!ht]
\begin{center}
\includegraphics[angle=-90,width=.47\linewidth]{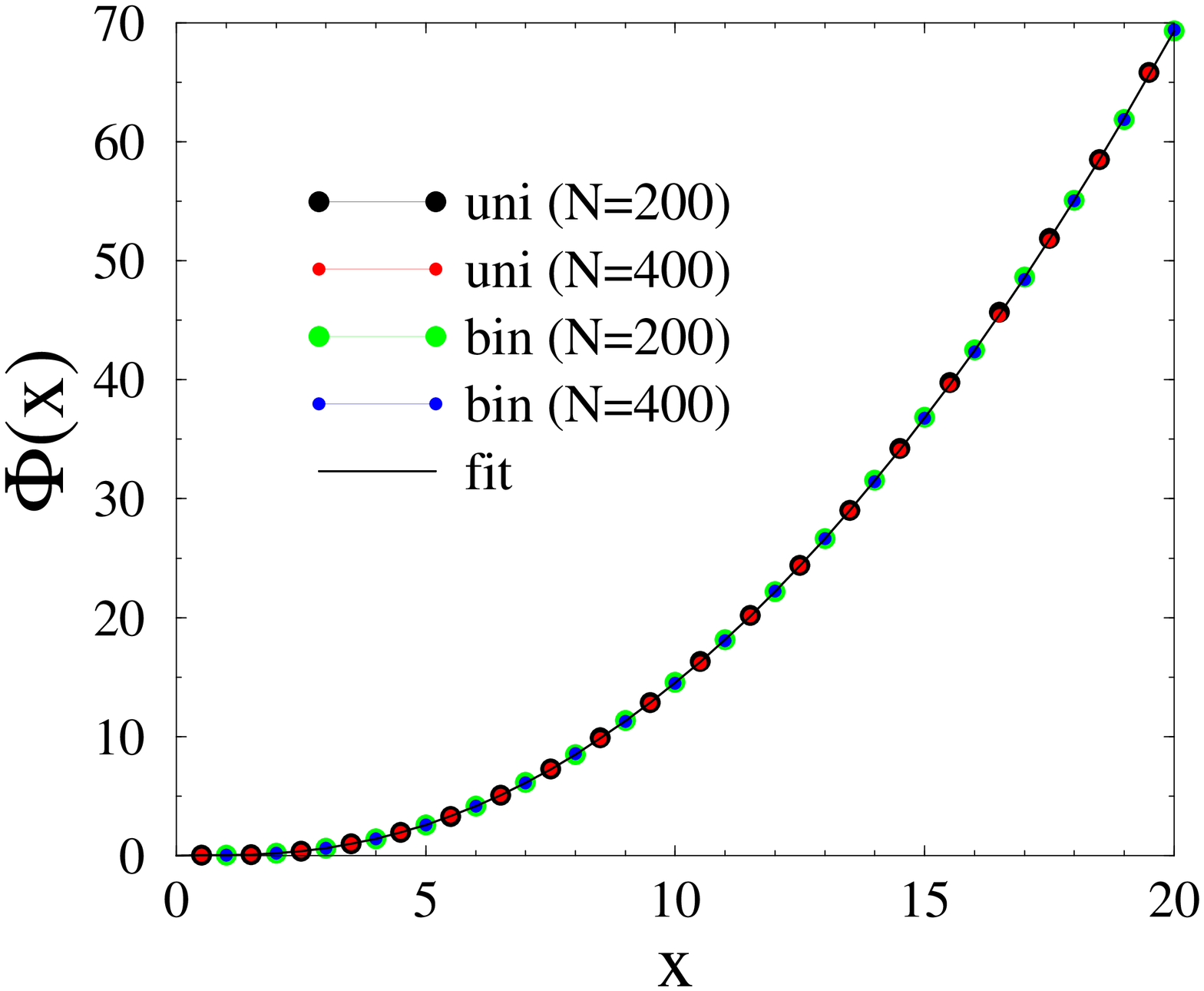}
\hskip 5pt
\includegraphics[angle=-90,width=.47\linewidth]{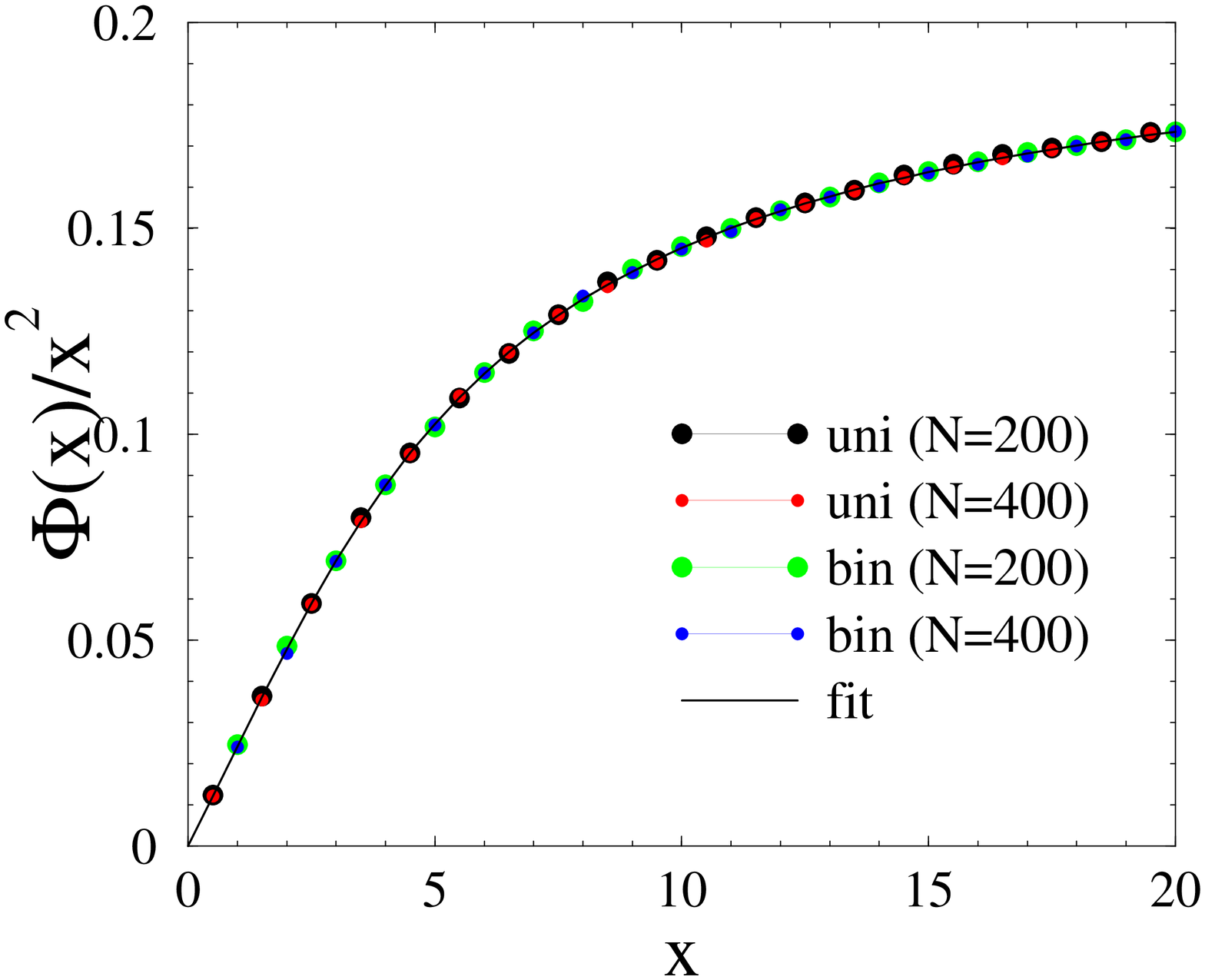}
\caption{\small
Finite-size scaling functions $\Phi(x)$ (left)
and $\Phi(x)/x^2$ (right), against $x=Nw^{2/3}$.
Symbols: data for uniform and binary distributions and $N=200$ and 400.
Full curves: common fit yielding~$A\approx0.205$.}
\label{fss}
\end{center}
\end{figure}

Figure~\ref{fss} shows plots of the finite-size scaling functions $\Phi(x)$ (left)
and $\Phi(x)/x^2$ (right), against $x=Nw^{2/3}$.
The right plot shows better the small-$x$ and large-$x$ regimes.
Data points correspond to uniform and binary distributions
and system sizes $N=200$ and 400.
Finite-size effects are negligible.
Full curves show a common three-parameter fit of the four data series
confirming the behavior~(\ref{phil}) at large $x$,
with $A\approx0.205$, in good agreement with~(\ref{a21}).

\section{Particle in a random potential: strong-disorder regime}
\label{strong}

This section is devoted to the strong-disorder regime ($w\to\infty$).
Our main goal is to explain the qualitatively different behavior of the mean return
probability $\bar{Q}$ for continuous and discrete distributions (see~figure~\ref{qave}).

\subsection{Continuous distribution of disorder}

We consider first the case where the on-site energies $V_a$
are drawn from a continuous distribution $f(V)$.
The differences between on-site energies at consecutive sites
are typically of the order of the disorder strength $w$, i.e., very large.
To leading order as $w\to\infty$, eigenstates are localized onto single sites:
site number~$a$ supports an eigenstate with amplitudes $\braket{a}{n}=\delta_{n,a}$
and energy $E_n=V_a$.
This is an example of the situation of `no equilibration' described in section~\ref{gal}.
We have
\beq
\bar{Q}\to1\qquad(w\to\infty).
\label{qinf}
\eeq

The corrections to this limit originate in the rare situations
where the ampli\-tudes~$\psi_a$ take appreciable values on more than one site.
This occurs when the differences between successive on-site energies $V_a$
are of order unity,
i.e., comparable to the free bandwidth, and therefore much smaller than~$w$.
The most probable of these rare disorder configurations
concern pairs of consecutive sites.
Assuming sites 1 and 2 form such a pair,
we are led to consider the effective problem
\beq
E\psi_a=\psi_{a+1}+\psi_{a-1}+V_a\psi_a
\eeq
($a=1,\,2$), with Dirichlet boundary conditions ($\psi_0=\psi_3=0$).
The neighboring on-site energies $V_0$ and $V_3$ are indeed generic,
and therefore very large.
This two-site problem boils down to the equations
\beq
\psi_2=(E-V_1)\psi_1,\qquad\psi_1=(E-V_2)\psi_2.
\eeq
We thus obtain two eigenstates, whose energies read
\beq
E_\pm=\frac{1}{2}\left(V_1+V_2\pm\sqrt{4+(V_1-V_2)^2}\right).
\eeq
The common value of the IPR of these eigenstates reads
\beq
I=1-\frac{2}{4+(V_1-V_2)^2}.
\eeq
By averaging this result over disorder, we obtain the estimate
\beq
\bar{Q}\approx1-4\;\bar{\left(4+(V_1-V_2)^2\right)^{-1}}.
\label{q1}
\eeq

In order to make this result more explicit, we set
\beq
V_a=w\,x_a.
\eeq
The reduced random variables $x_a$
have a symmetric continuous distribution $g(x)$ such that $\bar{x^2}=1$.
Changing variables in~(\ref{q1})
from $V_1$ and $V_2$ to their sum and difference,
we obtain after some algebra the first correction to the limit~(\ref{qinf}) in the form
\beq
\bar{Q}\approx1-\frac{2\pi\,C}{w},
\label{qstrong}
\eeq
where the constant
\beq
C=\int_{-\infty}^\infty g(x)^2\,\dd x
\eeq
measures the probability that two $x$ values are very near each other.

For the uniform distribution~(\ref{uni}), we obtain $C=1/(2\sqrt3)$, and therefore
\beq
\bar{Q}\approx1-\frac{\pi}{\sqrt{3}\,w}.
\label{qsuni}
\eeq
This estimate, shown in figure~\ref{qave}
as a straight line with slope $\pi/\sqrt{3}=1.813799$,
agrees with the data for the uniform distribution down to moderate values
of the disorder strength.

\subsection{Discrete distribution of disorder}

Let us now turn to the case where the on-site energies
are drawn from a discrete distribution,
considering for definiteness the binary distribution~(\ref{bin}).
In this situation, the mean return probability saturates
to the limit (see figure~\ref{qave})
\beq
\bar{Q}_\infty\approx0.373.
\label{qbin}
\eeq

The occurrence of such a non-trivial infinite-disorder limit
for the mean return probability
can be explained as follows.
In the strong-disorder regime,
one is naturally led to consider clusters of consecutive sites
with the same on-site potentials, either $+w$ or $-w$.
The chain is thus partitioned into molecules of various sizes,
along which on-site potentials are constant.
This line of reasoning dates back to an early work
on the spectra of disordered harmonic chains~\cite{domb}.
Since then it has been applied to a great deal of disordered systems~\cite{alea}.
For the symmetric binary distribution~(\ref{bin}), the size (number of sites)
$m$ of a molecule is geometrically distributed, with
\beq
p_m=\frac{1}{2^m}\qquad(m\ge1),
\label{pm}
\eeq
so that $\mean{m}=2$.

In a first approximation,
the eigenvalue equation~(\ref{hamv}) is to be solved separately on each molecule,
with Dirichlet boundary conditions.
A molecule of size $m$ possesses~$m$ eigenstates given by~(\ref{an}),
up to the replacement of the system size $N$ by the molecule size $m$.
We thus obtain the following molecular approximation for the infinite-disorder
mean return probability:
\beq
\bar{Q}_\mol=\frac{1}{\mean{m}}\sum_{m\ge1}p_m T_m,
\eeq
where $T_m$ is given by~(\ref{tfree}), up to the replacement of $N$ by $m$.
This yields
\beq
\bar{Q}_\mol=\frac{3}{2}-\ln 2-\frac{\ln 3}{4}=0.532199.
\eeq

The above prediction only provides an upper bound for $\bar{Q}_\infty$.
The reason is that degenerate molecular states of neighboring molecules may hybridize.
The resulting states are more delocalized that the original molecular ones,
and so $\bar{Q}_\infty$ is smaller than $\bar{Q}_\mol$.
The phenomenon of partial delocalization through hybridization
in the presence of a strong potential has been investigated on aperiodic chains~\cite{bl}.
To our knowledge, it has not been looked at in detail in the case of random disorder.

\begin{figure}[!ht]
\begin{center}
\includegraphics[angle=-90,width=.47\linewidth]{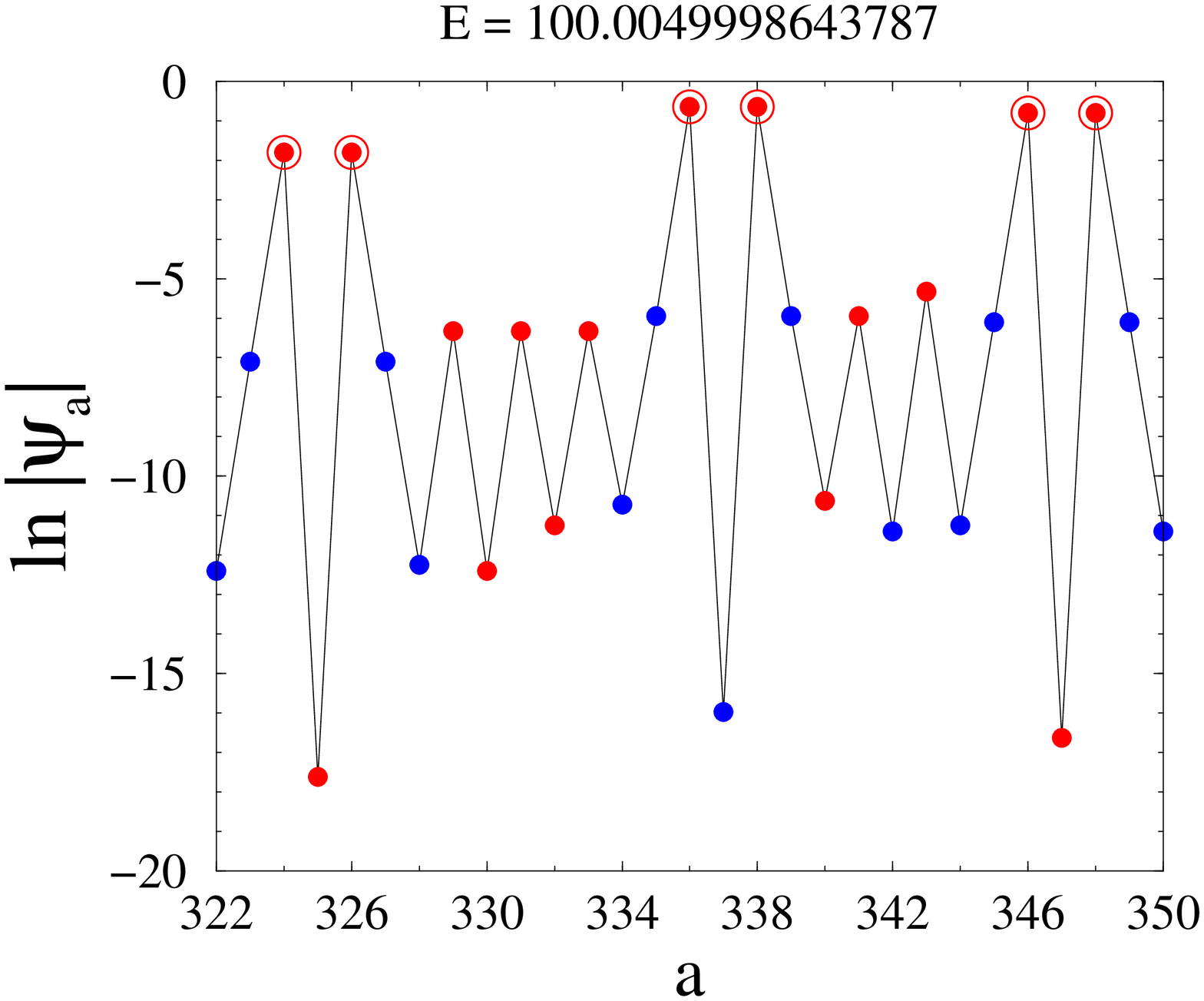}
\hskip 5pt
\includegraphics[angle=-90,width=.47\linewidth]{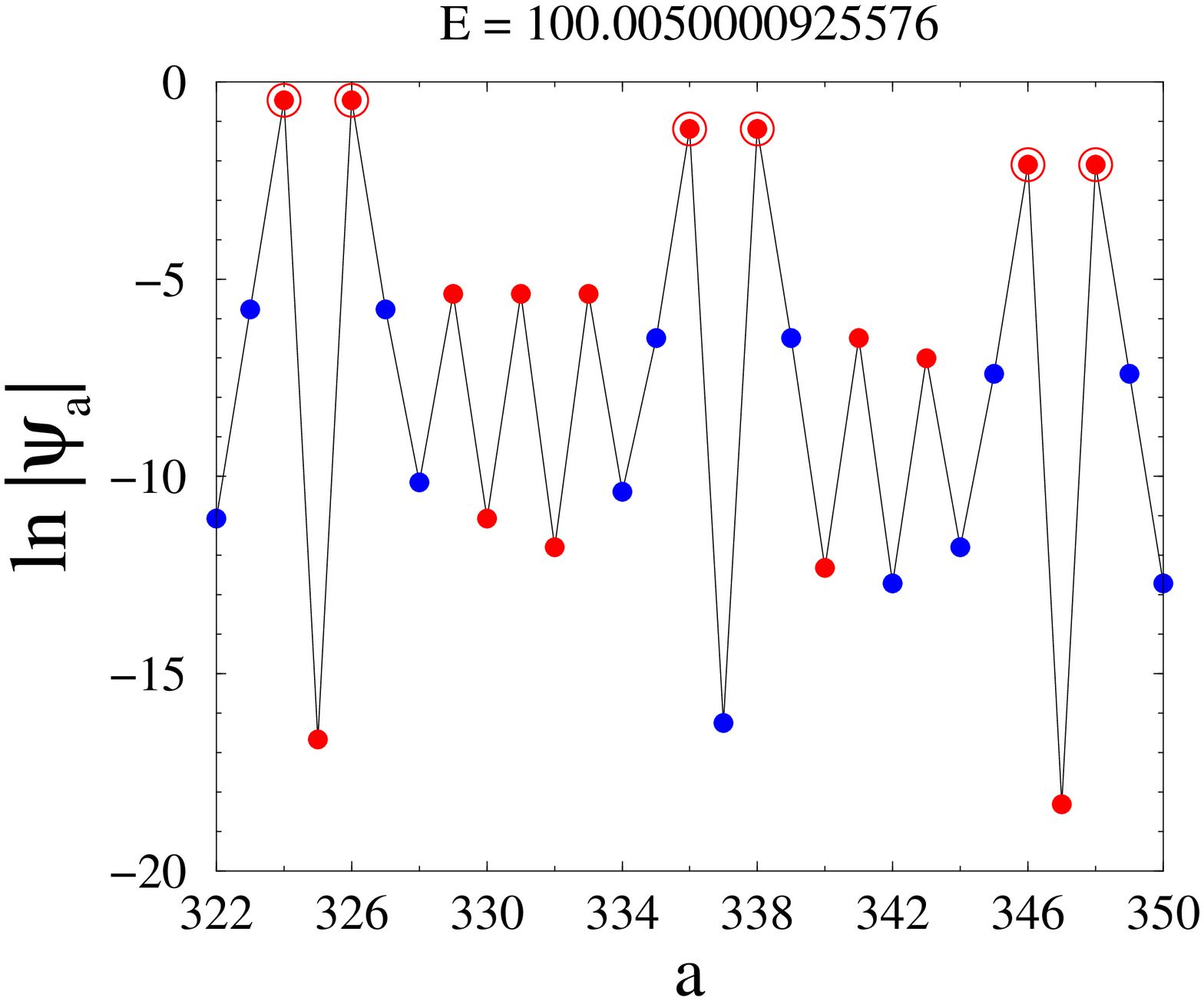}
\caption{\small
Two typical hybridized states in a chain of size $N=1000$
for a binary disorder with $w=100$.
Red (blue) symbols: $\ln\abs{\psi_a}$ at sites such that $V_a=w$ ($V_a=-w$).
Circled symbols: support of the states.}
\label{22}
\end{center}
\end{figure}

Figure~\ref{22} shows two typical hybridized states
whose energies are very close to each other.
The logarithm of $\abs{\psi_a}$ is plotted against position $a$ along the chain.
Circled symbols show the common {\it support} of both states,
i.e., the six sites where the amplitudes~$\psi_a$ are of order unity,
while those at all other sites fall off as inverse powers of $w$.

\begin{figure}[!ht]
\begin{center}
\includegraphics[angle=-90,width=.5\linewidth]{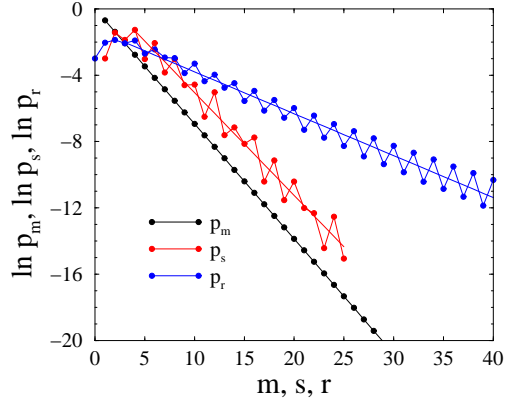}
\caption{\small
Logarithmic plots of the probability distributions
$p_m$ of the molecular size $m$
and $p_s$ and $p_r$ of the eigenstate size $s$ and range $r$
for a strong symmetric binary disorder.
Straight lines: least-square fits for $s$ and~$r$ larger than 3
yielding $\alpha_s\approx0.62$ and $\alpha_r\approx0.25$.}
\label{freqs}
\end{center}
\end{figure}

We now consider a generic eigenstate in the strong-disorder regime.
We define its {\it size}~$s$
as the number of sites where the amplitudes $\psi_a$ are of order unity
($s=6$ for the states shown in figure~\ref{22}).
We also define its {\it range} $r$
as the distance between the leftmost and the rightmost sites of its support
($r=348-324=24$ for the states shown in figure~\ref{22}).
Figure~\ref{freqs} shows logarithmic plots of the probability distributions~$p_s$
and~$p_r$ of eigenstate sizes and ranges.
A disorder strength $w=200$ already yields accurate asymptotic strong-disorder data.
The latter are gathered over $5\times10^6$ samples of size $N=200$.
The probability distribution $p_m$ of the molecular size~$m$
(see~(\ref{pm})) is also plotted for comparison.
The data suggest an exponential decay of all these distributions, of the form
\beq
p_m\sim\e^{-\alpha_m m},\qquad
p_s\sim\e^{-\alpha_s s},\qquad
p_r\sim\e^{-\alpha_r r}.
\eeq
For the molecular size $m$,~(\ref{pm}) yields $\alpha_m=\ln 2=0.693147$.
For the eigenstate size~$s$ and range $r$,
the least-square fits shown in figure~\ref{freqs} yield
$\alpha_s\approx0.62$ and $\alpha_r\approx0.25$.

The appreciable differences between $\alpha_r$ and $\alpha_m$
and between $\bar{Q}_\infty$ and $\bar{Q}_\mol$
testify that the strong-disorder hybridization mechanism described above
has sizeable consequences in generic circumstances.

\section{Discussion}
\label{disc}

The present work has been devoted to equilibration in small isolated quantum systems.
In order to deal with all possible initial states on the same footing
-- assuming a preferential basis has been chosen once for all --
we have been led to consider the full matrix $Q_{ab}$
of asymptotic transition probabilities in the chosen basis,
and especially its trace, $T=\tr Q$,
which characterizes the degree of equilibration of the system
launched from a typical initial state, from the viewpoint of the chosen basis.

This approach has been substantiated
by means of an in-depth study of a simple one-body problem,
namely a tight-binding particle on a finite chain of~$N$ sites.
In the regime of free propagation (section~\ref{free}),
the trace $T$ saturates to the limit $T=3/2$,
a finite factor above the minimal value $T_\min=1$,
testifying good equilibration.
In the presence of a random potential whose disorder strength is $w$,
the trace grows asymptotically as $T\approx\bar{Q}N$ in the Anderson localized regime.
This linear growth is proportional to the maximal value $T_\max=N$,
corresponding to no equilibration (i.e., full localization).
The amplitude $\bar{Q}<1$ is the mean return probability of the particle
to its starting point.
Its dependence on the type and strength of disorder has been studied in detail.
In the weak-disorder situation,
we have evidenced the universal power law
$\bar{Q}\approx A\,w^{4/3}$ in the localized regime, with $A\approx0.21$,
as well as the finite-size scaling law~(\ref{tfss}) interpolating between the
ballistic and the weak-disorder localized regimes.
Both features are intimately related to the anomalous band-edge scaling
characterizing the most localized energy eigenstates.

The body of this paper deals with a single example of a simple quantum system,
i.e., a tight-binding 1D particle with or without a random potential.
It is tempting to investigate whether the idea of the present approach,
i.e., considering the matrix~$Q_{ab}$ of asymptotic transition probabilities,
and especially its trace $T$,
could bring valuable information in other situations.
The present construction is certainly costly,
as it requires the knowledge of all energy eigenvalues and eigenvectors.
Its scope is therefore a priori limited to small systems.
The latter limitation is however not inconsistent with the initial purpose of the approach,
which is precisely to quantify the degree of equilibration
-- or of lack of equilibration -- of {\it small} isolated quantum systems.

Finally, as far as many-body quantum systems are concerned,
the simplest situation one may have in mind is a chain of spins $s=1/2$.
The dimension $N=2^M$ of the Hilbert space grows exponentially fast
with the number~$M$ of spins.
The trace~$T$ may therefore a priori vary over an exponentially large range.
In particular, one can speculate that an exponential growth of $T$
with the chain length $M$
might be an alternative signature of many-body localization.
Besides direct numerical diagonalizations,
analytical approaches using integrability and free-fermion techniques
might also allow some progress in this direction.

\ack

It is a pleasure to thank R Balian, G Ithier, P Krapivsky,
K Mallick, G Misguich and V Pasquier for useful discussions.

\section*{References}

\bibliography{revised.bib}

\end{document}